\documentclass[prb,twocolumn,amsmath,superscriptaddress]{revtex4}
\usepackage{graphicx}

\usepackage{bm}

\begin{document}

\title{Non exponential spin relaxation in magnetic field in quantum
wells with random spin-orbit coupling}
\author{M.M. Glazov}
\affiliation{A. F. Ioffe Physico-Technical Institute, Russian
Academy of Sciences, 194021 St. Petersburg, Russia}
\author{E.Ya. Sherman}
\affiliation{Department of Physics and Institute for Optical Sciences, University of Toronto, 60 St.
George Street, Toronto, Ontario, Canada M5S 1A7}

\begin{abstract}
We investigate the spin dynamics of electrons in quantum wells
where the Rashba type of spin-orbit coupling is present in the
form of random nanosize domains. We study the effect of magnetic
field on the spin relaxation in these systems and show that the
spatial randomness of spin-orbit coupling limits the minimum
relaxation rate and leads to a Gaussian time-decay of spin
polarization due to memory effects. In this case the relaxation
becomes faster with increase of the magnetic field in contrast to the
well known magnetic field suppression of spin relaxation.
\end{abstract}

\maketitle

The effect of magnetic field on spin relaxation in low-dimensional
structures in the presence of spin-orbit (SO) coupling is an
interesting theoretical and experimental problem. The
understanding of this effect which allows to a certain extent the
engineering of the spin dynamics in these systems can have a
crucial impact on the possible spintronics
applications.\cite{Burkard99} The SO coupling of electrons in
two-dimensional (2D) zincblende-based systems with the structural
asymmetry grown along the [001] direction is described by the
Rashba Hamiltonian $\hat{H}_{R}=\alpha(\hat
\sigma_{x}{k}_{y}-\hat\sigma_{y}{k}_{x})$ (Ref.
[\onlinecite{Rashba84}]), where $\alpha$ is the coupling constant,
$\hat{\sigma}_i$ are the Pauli matrices, and ${\bm k}$ is the
in-plane wavevector of the electron. The random spin precession
necessary for the spin relaxation is introduced by electron
scattering by impurities, phonons, other
electrons\cite{gi_jetp,Weng03} and due to random dynamics in
regular systems.\cite{Pershin04} In the "dirty" limit, $\alpha
k\tau/\hbar\ll 1$ with $\tau$ being the momentum relaxation time,
the motional narrowing leads to the Dyakonov-Perel'
mechanism\cite{Dyakonov72} causing the exponential spin relaxation
with the rate of the order of $(\alpha k/\hbar)^{2}\tau$. The spin
relaxation rate decreases when a magnetic field is applied
\cite{Ivchenko73} due to the effect of the Lorenz force on the
orbital movement of the electron. From the analysis of the spin relaxation
in a magnetic field valuable information both on spin and charge
dynamics can be extracted.\cite{Glazov04} In the clean limit,
$\alpha k\tau/\hbar\gg 1$ spin relaxation, on a time scale of the
order of $\tau$, occurs on the top of the spin precession
\cite{Gridnev01}. Various mechanisms of spin relaxation in solids
are reviewed in Ref. [\onlinecite{Zutic04}].

In most quantum wells (QW) the Rashba-type SO coupling is achieved
by asymmetric remote doping at the sides of the well; the same
doping forms a smooth random potential scattering electrons. A
crucially important step in being able to quickly manipulate the
spins has been made by demonstrating that, by applying external
bias across the quantum well, it is possible to change the
magnitude of $\alpha$
\cite{Knap96,Nitta97,Koga02,Miller03,Harley03}. Almost all
previous studies assumed the $\alpha$  parameter to be constant in
space. However, the evidence is growing that the SO coupling both
in zincblende and Si$_{x}$Ge$_{1-x}$/Si-based QWs
\cite{Sherman03,Golub04} is a random function of
coordinate. The randomness arises due to imperfections in the
system, e.g. either due to shot noise accompanying the doping
\cite{Sherman03} or due to random variations of the
bonds at Si/Ge interfaces \cite{Golub04} in the
Si$_{x}$Ge$_{1-x}$/Si systems considered as a hope for spintronics
due to a very small SO coupling
there.\cite{Jantsch02,Tahan04} Therefore, the spin of a
moving electron interacts with a randomly time-dependent effective
SO field. This  randomness leads to an irregular precession of
spins and contributes to the spin relaxation rate. Here we
concentrate on the systems where the Rashba term dominates in the
SO-coupling, like InGaAs-based structures \cite{Nitta97}, and
demonstrate new qualitative effects of the randomness on the spin
relaxation in a magnetic field such as a possibility of the {\it
non-exponential} spin relaxation, and an {\it increase} of the spin
relaxation rate in a strong magnetic field. The effect of the
Dresselhaus coupling arising due to the unit cell asymmetry
\cite{Dresselhaus} will be briefly discussed in the context of our
results also. The importance of the randomness of electric field
of dopants causing the SO coupling on the donor electrons was
first demonstrated  by Mel'nikov and Rashba \cite{Melnikov72} for
bulk Si. Recently, Vagner {\it et al.} \cite{Vagner98} showed that
a random SO coupling for electrons in GaAs mesoscopic rings
interacting with nuclear spins can lead to an Aharonov-Bohm-like
effect.

Since we are interested in the description of random spin and
charge dynamics, we consider as an example a QW of the width  $w$
extended between $-w/2<z<w/2$, and surrounded by two
$\delta$-doped layers at $z=\pm z_{0}$, $z_0 \gg w/2$, with a mean
two-dimensional concentration of dopants $N_{u}$ (upper layer) and
$N_{d}$ (lower layer) with charge $\left|e\right|$. The local SO
coupling, being a linear response of the system to the
symmetry-breaking perturbation, is presented as
$\alpha_{\mathrm{R}}({\bm\rho})=\alpha_{\rm SO}|e|{E}_{z}({\bm\rho})$, where ${E}_{z}({\bm\rho})$ is the $z-$
component of electric field of the dopant ions, ${\bm \rho }$ is
the 2D coordinate at the $z=0$ plane, and $\alpha_{\rm SO}$ is a
system-dependent phenomenological parameter. The $z$-and in-plane
$i=x,y$ components of the Coulomb field of the dopant ions are
given by \cite{Ando}
\begin{eqnarray}
{E}_{z}({\bm \rho}) &=&-\frac{\left|{e}\right|}{\epsilon}
\sum_{_{j}}\frac{z_{j}} {[({\bm
\rho}-\bm{r}_{j\Vert})^{2}+z_{0}^{2}]^{3/2}},
\nonumber\\
{E}_{i}({\bm \rho }) &=&\frac{\left| {e}\right| }{\epsilon}
\sum_{_{j}}\int\limits_{0}^{\infty }qA_{q}
\frac{\partial}{\partial \rho_i}
J_{0}\left(q\left|{\bm\rho}-\bm{r}_{j\Vert}\right|\right)dq,
\end{eqnarray}
where $\bm{r}_{j}=(\bm{r}_{j\Vert },z_{j})$,
$\bm{r}_{j\Vert}=(r_{x},r_{y})$ is the 2D radius-vector of the
$j$th dopant ion, $z_{j}=\pm z_{0}$ for the upper and lower layer,
respectively, $\epsilon $ is the dielectric constant,
$J_{0}(q\rho)$ is the zero-order Bessel function, and
$A_{q}=\exp(-qz_{0})/(q+q_{s}),$ where $q_{s}=\left(2\pi
e^{2}/\epsilon\right)dN/d\varepsilon_{F}$ describes the
Thomas-Fermi screening of the in-plane Coulomb field with
$N=N_{u}+N_{d}$ being the electron concentration, and
$\varepsilon_{F}$ being the Fermi energy. As a result, electrons
move in a smooth 2D random potential $U({\bm \rho})$ with ${\bm
E}_{\parallel}({\bm \rho})=-\nabla_{\parallel}{U}({\bm \rho})$,
with the mean value $\langle U\rangle=0$.

We consider the semiclassical movement of the electron in external
magnetic field $\bm{H}$:
\begin{equation}\label{newton}
\hbar\frac{d\bm{k}}{dt}= e{\bm E}_{\parallel}({\bm \rho})+
\frac{\hbar e}{mc}\: \bm{k}\times\bm{H},
\end{equation}
where $m$ is the electron effective mass. The momentum relaxation
time for the electron moving in a smooth random potential is given
by\cite{mirlin00}
\begin{equation}\label{tau:approx}
\frac{1}{\tau}=-\frac{1}{m^2v_F^3}
\int\limits_{0}^{\infty}\frac{d\rho}{\rho}
\frac{dC_{UU}(\rho)}{d\rho},
\end{equation}
where $C_{UU}(\rho)=\langle U({\bm \rho}) U(0)\rangle$ is the
correlation function of the random potential, and $v_F=\hbar
k_F/m$ is the Fermi velocity corresponding to the Fermi wavevector
$k_F=\sqrt{2\pi N}$. The relaxation time can be estimated as
$\tau\sim\tau_{d}\varepsilon_{F}^{2}/\left\langle
U^{2}\right\rangle$, the potential fluctuation $\langle
U^{2}\rangle\sim N\left(e^{2}/\epsilon q_{s}z_{0}\right)^{2}$, and
$\tau_{d}$ is the time of passing through one domain of random
force and random SO coupling, with typical size $z_{0}$ in this
case. In InGaAs based structures with $m=0.02m_0$, where $m_0$ is
the bare electron mass, $q_s$ is close to $2\times10^6$ cm$^{-1}$.
For numerical estimates we use the parameters from Refs.
[\onlinecite{Koga02,Zeta}] with $N_{u}=10^{11}$ cm$^{-2}$ and
$N_{d}=3\times 10^{11}$ cm$^{-2}$. Then the ratio
$\varepsilon_{F}^{2}/\langle U^{2}\rangle$ is of the order of 100
leading to $\tau$ two orders of magnitude longer than $\tau_d\sim
0.03$ ps. In weak fields $\omega_c\tau\ll 1$, where
$\omega_c=|e|H/mc$ is the cyclotron frequency, the electron  path
is random, while at $\omega_c\tau>1$ it becomes close to the
regular circular movement in magnetic field with a random path of
the orbit center. At $m=0.02m_0$ the crossover from the random to
the regular movement occurs at $H\sim 0.1$ T. The semiclassical
description of the orbital movement is possible in a non-quantizing
field up to $H\sim 1$ T, where $\hbar\omega_c/\varepsilon_{F}\sim
0.1$.

The Rashba parameter is the sum of the mean value
$\langle\alpha\rangle=2\pi\alpha_{\rm SO}{e^2}(N_{d}-N_{u})/\epsilon$, and a
random term with the zero mean contribution: $\alpha({\bm
\rho})=\langle\alpha\rangle+\delta\alpha({\bm \rho })$ and the
variation \cite{Efros89}
\begin{equation}
\langle\alpha ^{2}\rangle-\langle\alpha\rangle^{2}=
\langle\alpha\rangle^{2}
\frac{N}{8\pi\left(N_{d}-N_{u}\right)^{2}z_{0}^{2}}.
\end{equation}
The random and regular contributions become comparable at
$Nz_{0}^{2}\lesssim 0.1 $; the condition satisfied in most of the
experimentally investigated QWs. The regular term
$\langle\alpha\rangle$ can be reduced or completely removed by
applying an electric bias across the quantum well, but even in
this case the random term $\delta\alpha({\bm \rho})$ remains and
causes spin relaxation. Figure 1 presents a pattern of
$\alpha_{\mathrm{R}}({\bm\rho})$ obtained by a Monte-Carlo
produced white-noise distribution of dopant ions corresponding to
the experimental data of Ref.~[\onlinecite{Koga02}]. As can be
seen in Fig. 1, the random variations of
$\alpha_{\mathrm{R}}\left({\bm \rho}\right)$ are large and
correlated on the distances of the order of $z_{0}$. Therefore,
the pattern of SO coupling is random nanosize domains rather than
a constant.

\begin{figure}[htb]
\begin{center}
\includegraphics[width=6.0cm]{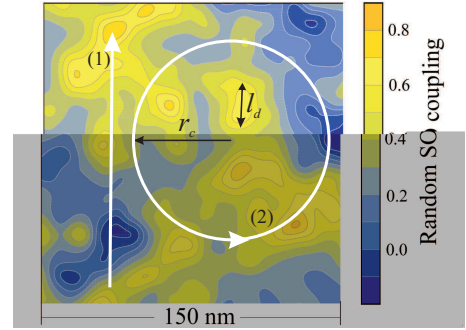}
\end{center}
\caption{(Color online) A realization of the random Rashba SO
coupling (in arbitrary units) due to fluctuations of the dopant
concentration, $N_{u}=10^{11}$ cm$^{-2}$, $N_{d}=3\times 10^{11}$
cm$^{-2}$, and $z_{0}=12$ nm. Curves (1) and (2) illustrate
possible electron trajectories without magnetic field and in the
field in the ballistic regime, $r_c$ is the cyclotron radius and
$l_d$ is the domain size (see Eq. (\ref{alpha:corr})).}
\label{fig:alphas}
\end{figure}

In general for a coordinate-dependent $\alpha\left({\bm
\rho}\right)$, the Rashba Hamiltonian should be symmetrized to
have the Hermitian form but we assume semiclassical movement of
electrons  and describe the spin dynamics with a momentum and
coordinate dependent effective Larmor frequency ${\bm
\Omega}\left[{\bm
k}(t),{\bm\rho}(t)\right]=2\alpha({\bm\rho})(k_y,-k_x)/\hbar$.

The semiclassical spin dynamics of electron neglecting the Larmor
contribution of the external magnetic field \cite{Larmor} to the
spin precession is described by:
\begin{equation}\label{dynamics}
\frac{\partial {\bm s}}{\partial t}+ {\bm s}\times{\bm
\Omega}\left[{\bm k}(t),{\bm \rho}(t)\right]=0.
\end{equation}
Equation (\ref{dynamics}) means that the electron spin precesses in the
effective field caused by the coordinate and wavevector dependent
spin-splitting that fluctuates due to the variations of the
wavevector and electron position. It is well known that the spin
dynamics is governed by the fluctuations of the Larmor frequency
$\bm\Omega$ provided that their timescale is shorter than the spin
relaxation time. Let us introduce the autocorrelation function
$C_{\Omega\Omega}(t) = \langle {\bm \Omega} \left[{\bm k}(t),{\bm
\rho}(t)\right] {\bm \Omega} \left[({\bm k}(0),{\bm
\rho}(0)\right] \rangle$ which is taken along the semiclassical
path $\bm \rho(t)$ obtained by integrating Eq. (\ref{newton}).
Further we restrict ourself to the $\tau$-approximation
\cite{Bronold04,Sinitsyn05} with $\tau$ given by Eq.
(\ref{tau:approx}). It allows us to present the solution of Eq.
(\ref{dynamics}) for the ensemble mean value of spin $\langle s_z
(t)\rangle$ as\cite{semenov03}
\begin{equation}\label{sz:time:dep}
\langle s_{z}(t)\rangle=\langle s_{z}(0)\rangle \exp \left[
-\int\limits_{0}^{t}dt^{\prime}\int\limits_{0}^{t^{\prime}}dt^{\prime\prime}
C_{\Omega\Omega}(t'-t'') \right].
\end{equation}

In high-mobility structures the smooth random potential of dopant
ions causes two main sources of randomness leading to a spin
relaxation: (i) random $\alpha\left({\bm \rho}\right)$ due to
fluctuations of the electric field ${E}_{z}$ and (ii) random
direction of $\bm{k}$. The former contribution causes the
randomness in the amplitude of the SO field and, in turn, the
precession rate, on the time scale $\tau_{d}$, while the latter
causes the random orientation of the precession axis on the much
longer time scale $\tau$. These effects determine the correlator
in Eq.(\ref{sz:time:dep}).

To take into account different possible mechanisms of randomness
of the SO coupling (fluctuations in the dopant concentrations,
randomness of the bonds at the interfaces, variations of the
directions of the growth axis, random strain, etc.) we consider a
simple model with
\begin{equation}\label{alpha:corr}
\langle\delta\alpha({\bm\rho})\delta\alpha(0)\rangle=
\langle(\delta\alpha)^2\rangle e^{-\rho/l_d},
\end{equation}
where $l_d=\tau_d v_F$ is the characteristic domain size of the
order of 10 nm. The details of the shape of the correlator in
Eq.(\ref{alpha:corr}) have no qualitative influence on our
results. One can show that the correlator of the effective Larmor
frequencies in Eq.(\ref{sz:time:dep}) is given by
\begin{eqnarray}\label{omega:corr}
&& C_{\Omega\Omega}(t) = \frac{4k^2}{\hbar^2} C_{kk}(t) \times
\\ \nonumber
&&\left[ \langle\alpha\rangle^2 + \langle(\delta\alpha)^2\rangle
\exp \left( -\frac{2}{\omega_c\tau_d}
\left|\sin\frac{\omega_{c}t}{2}\right| - \frac{\delta r}{l_d}
\right) \right]
\end{eqnarray}
Here $\delta r$ is the displacement of the center of cyclotron
orbit due to drift in the random potential,\cite{mirlin00,perel}
and $C_{kk}(t)=\cos\omega_c t\: e^{-t/\tau}$ is the electron
momentum correlator in the magnetic field. The first term in the
square brackets gives the contribution to the spin dynamics due to
regular SO coupling, the second one stands for the contribution
from random fluctuations of $\alpha$. In Eq. (\ref{omega:corr})
the correlations between the fluctuations of SO coupling and
momentum are neglected since they have drastically different time
scales $\tau_d$ and $\tau$, respectively.

To begin we consider the case of a not very strong field
$\omega_c \tau_d \ll 1$, where $\delta r\gg l_d$. In the collision
dominated regime $\langle \alpha \rangle k \tau/\hbar\ll 1$, spin
decays exponentially and the relaxation rate obtained with Eqs.
(\ref{sz:time:dep}) and (\ref{alpha:corr}) is given by
$\Gamma_{zz}=\Gamma_{zz}^{(c)}+\Gamma_{zz}^{(r)}$, where
\begin{equation}\label{gamma:zz:c}
\Gamma_{zz}^{(c)}=\frac{4\langle(\delta\alpha)^2\rangle
k^2}{\hbar^2} \tau_d,\quad
\Gamma_{zz}^{(r)}=\frac{4\langle\alpha\rangle^2 k^2\tau}
{\hbar^2\left(1+\omega_c^2\tau^2\right)}.
\end{equation}
At $H=0$ both terms are the product of the typical spin-splitting
squared and correlation time corresponding to the D'yakonov-Perel'
relaxation mechanism.\cite{Dyakonov72} Here the electron passes
many domains moving along the straight path (line (1) in Fig.
\ref{fig:alphas}), the spin splitting in each domain is different
and the time scale of the variations of the effective Larmor field
$\bm \Omega\left[\bm k(t),\bm \rho(t)\right]$ is $\tau_d$. In
a magnetic field at $\omega_c^2\tau^2\gg 1$, in the case of the
regular SO coupling, the effective Larmor frequency $\bm
\Omega[\bm k(t+T/2)]=-\bm \Omega[\bm k(t)]$ is reversed each half
of the cyclotron period on the circular trajectory causing spin
precession backwards, and the relaxation slows down. It is not the
case when the SO constant fluctuates in space and
$\langle\alpha\rangle=0$: if the cyclotron radius is large enough
($\omega_c\tau_d \ll 1$), the random contributions to ${\bm
\Omega}\left[{\bm k}(t),{\bm\rho}(t)\right]$ are not correlated,
the reversal of precession does not take place making the
relaxation rate $H$-independent. We emphasize that the source of
randomness here is not the momentum scattering but the random
fluctuations of the spin-orbit constant.  A numerical estimate at
$\sqrt{\langle(\delta\alpha)^2\rangle}=1.5\times 10^{-10}$ eVcm
and experimental conditions of Ref.[\onlinecite{Koga02}] gives
$1/\Gamma_{zz}^{(c)}\sim 60$ ps [\onlinecite{GaAs}].

To understand the qualitatively different situation when the
magnetic field becomes so strong that after a cyclotron period
electron returns to the same domain of spin-splitting, we consider
the 'compensated' case with $\langle \alpha\rangle=0$. For the
particle in a smooth potential the condition for such a ballistic
regime is $\langle\left|{\bm \rho}(t+T)-{\bm
\rho}(t)\right|\rangle = \delta r\ll l_{d}$, $T=2\pi/\omega_c$,
which can be achieved at $\omega_c\tau\gg
(\tau/\tau_d)^{2/3}$.\cite{mirlin00,perel} This condition can be
fulfilled in non-quantizing fields at $\tau \gg 10$ ps,
corresponding to high mobilities $\mu\gtrsim 10^{6}$
cm$^{2}$/(Vs). In such a case on each cyclotron period the
electron passes exactly the same configuration $\delta \alpha(\bm
\rho)$ (see circle (2) in Fig.~\ref{fig:alphas}). It means that
the effective correlation time of Larmor frequency would be much
longer than $\tau_d$ and the spin relaxation becomes {\it faster}
in a strong magnetic field. This result is in a sharp contrast to
the well-known magnetic field suppression of the spin
relaxation\cite{Ivchenko73} in Eq.(\ref{gamma:zz:c}). During the
first cyclotron revolution the spin dynamics is the same as in the
absence of a magnetic field. For subsequent revolutions the spin
rotates by the same angle as during the first one because the
electron passes the same configuration of $\alpha(\bm \rho)$. This
memory effect enhances the spin relaxation rate on each cycle and
makes the spin relaxation non-exponential, namely
\begin{equation}\label{nonexp}
\frac{\langle s_z(t) \rangle}{\langle s_z(0) \rangle} =\exp
\left\{-\Gamma_{zz}^{(c)}\left[K^2T+(1+2K)(t-KT) \right] \right\},
\end{equation}
with $K=\lfloor t/T \rfloor$ and $\lfloor\ldots \rfloor$ standing
for the integer part. It is interesting to note that Cremers {\it
et al.} \cite{Cremers03} found that the effect of non-uniform SO
coupling on weak localization in quantum dots can be interpreted
as a speeding up of spin relaxation. The cusps in $\langle
s_z(t)\rangle$ at $t=nT$ are smoothed out if one takes into
account the drift of the cyclotron orbit $\delta r$ in the random
potential. At $t\gg T$ we obtain $\langle s_z(t) \rangle \sim \exp
{(-\Gamma_{zz}^{(c)}t^2/T)}$. Figure 2 shows the calculated time
dependence of the $\langle s_z\rangle$ for different values of the
external magnetic field.

\begin{figure}[tbh]
\begin{center}
\includegraphics[width=5.0cm]{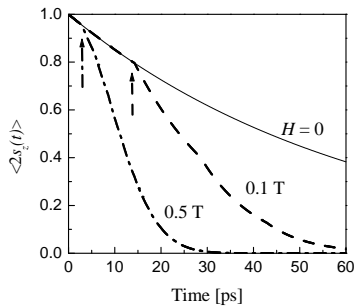}
\end{center}\vspace{-5mm}
\caption{Time dependence of $\langle s_{z}\rangle$ in the
compensated  ballistic regime without magnetic field (solid),
$H=0.1$~T (dash) and $H=0.5$~T (dash-dot). The arrows show the first
cyclotron revolution. The parameters of calculation were as
follows: $m=0.04m_0$,
$\sqrt{\langle(\delta\alpha)^2\rangle}=1.5\times 10^{-10}$~eVcm,
$l_d=12$~nm, $k_F = 1.6\times 10^6$~cm$^{-1}$, and $\langle
s_{z}(0)\rangle=1/2.$ This set of parameters can be realized in
the structure investigated in Ref.[\onlinecite{Koga02}] at
$N_{a}=N_{d}=2\times10^{11}$ cm$^{-2}$.} \label{fig:plane}
\end{figure}

We note that in the high-mobility structures the collision
dominated regime can be violated and spin relaxation occurs on the
top of spin precession making Eq. (\ref{sz:time:dep})
invalid.\cite{Gridnev01} However in strong magnetic field
$\omega_c \gtrsim \langle\alpha k/\hbar\rangle$ the applicability
of Eq. (\ref{sz:time:dep}) can be justified, besides, the regular
contribution to the spin relaxation is strongly suppressed by the
magnetic field. Therefore the spin relaxation rate drops to its
residual value of the order of $\Gamma_{zz}^{(c)}$ solely
determined by the fluctuations of random SO-coupling. The further
increase of magnetic field to
$\omega_c\tau\sim(\tau/\tau_d)^{2/3}$ makes the spin relaxation
non-exponential (see Eq.(\ref{nonexp})) with the characteristic
decay time $\tau_s\sim\sqrt{T/\Gamma_{zz}^{(c)}}$. Finally, we
mention that the addition of the regular Dresselhaus SO coupling
will not change our main results since it is suppressed by the
magnetic field in the same way as the Rashba contribution, and,
therefore, only the effect of a random Rashba coupling remains in
a sufficiently strong magnetic field.

To conclude, we have shown that the inevitable randomness of SO
coupling has an important effect on the spin dynamics in 2D
structures, e.g. it determines the minimum spin relaxation rate
achieved in a magnetic field.  In the ballistic regime, achieved in
sufficiently strong magnetic fields, the domain pattern of SO
coupling leads to the non-exponential decay of spin polarization being
a manifestation of a spin memory effect. The spin relaxation in a
system with random SO-coupling becomes faster with the increase of
the magnetic field.\cite{Ostreich04} These effects limit
possibilities of spin manipulation. The predicted magnetic field
dependence of the spin relaxation can be observable in 2D
structures\cite{Lewis02} with $\mu\gtrsim 10^{6}$ cm$^{2}$/(Vs),
where the "ballistic" regime is achievable in non-quantizing
fields.

M.M.G. acknowledges the financial support by RFBR and the "Dynasty"
foundation -- ICFPM. E.Ya.S. is supported by the FWF P15520 grant
and the DARPA SpinS program. We are grateful to E.L. Ivchenko,
J.E. Sipe, J. Sinova, and V.L. Pokrovsky for very valuable
discussions and suggestions, and to P. Marsden for the critical
reading of the manuscript.


\begin{thebibliography}{99}
\bibitem{Larmor} In zincblende structures the contribution of the magnetic
                 field at $H=1$ T in the spin precession rate is an order
                 of magnitude smaller than the contribution of the SO term.
\bibitem{GaAs}  This value is typical for InGaAs-based structures. For symmetric GaAs
                            QWs, where $\alpha_{\rm SO}$ is considerably smaller, the fluctuations of
                the electic field of the dopant ions shall lead to much longer spin relaxation times.
\end{thebibliography}
\end{document}